\documentclass[prd,showpacs,preprintnumbers,amsmath,amssymb]{revtex4}
\usepackage{graphicx}
\usepackage{dcolumn}
\usepackage{bm}
\usepackage{latexsym}
\usepackage{amsfonts}
\usepackage{amssymb}
\usepackage{amsmath}

\begin{document}

\preprint{KUNS-1963}

\title{ Low Energy Effective Action for Ho\v rava-Witten Cosmology }

\author{Sugumi Kanno}
\email{sugumi@tap.scphys.kyoto-u.ac.jp}
\affiliation{
 Department of Physics,  Kyoto University, Kyoto 606-8501, Japan
}%

\date{\today}

\begin{abstract}
As a supersymmetric extension of the Randall-Sundrum
model, we consider a 5-dimensional Ho\v rava-Witten type theory,
and derive its low energy effective action.
The model we consider is a two-brane system with a bulk
 scalar field satisfying the BPS condition.
We solve the bulk equations of motion using a gradient expansion method,
and substitute
the solution into the original action to get the 4-dimensional
effective action. The resultant effective theory can be casted into
the form of Einstein gravity coupled with two scalar fields,
one arising from the radion, the degree of freedom of the inter-brane
distance, and the other from the bulk scalar field.
We also clarify the relation between our analysis
and the moduli approximation. 
\end{abstract}

\maketitle

\section{Introduction}

Nowadays, it is believed that superstring theory is the most 
promising candidate for quantum theory of gravity. 
Remarkably, it can be consistently formulated only in 10 dimensions
~\cite{Polchinski}. 
This fact requires a mechanism to fill the gap between our real world and
the higher dimensions. Conventionally, the extra dimensions are considered
to be compactified to a small compact space of the Planck scale.
However, recent developments in superstring theory has lead to 
a new idea, the so-called braneworld.
 The braneworld scenario has been a subject of intensive investigations
 for the past few years~\cite{Maartens:2003tw}. The simplest toy
model of the braneworld is constructed by Randall and Sundrum in which
the bulk is empty~\cite{RS}. However, in superstring theory, scalar 
fields are ubiquitous. Indeed, generically a dilaton and moduli exist in
the bulk because they correspond to modes associated with a 
closed string. Thus, including a bulk scalar field is more natural 
from the string theory point of view.

In addition to the above stingy point of view, a model 
with a bulk scalar field is interesting from the cosmological point of view
as well. For example, a bulk scalar field can drive inflation on the brane
 although the bulk spacetime never inflate~\cite{bulk}. 
This kind of inflationary scenario certainly deserves further investigations. 

In the presence of a bulk scalar field, the physics of 
branes is complicated. However when both the bulk potential and 
the effective brane tension are related and under control,
 the effective 4-dimensional cosmological constant vanishes. 
This is the so-called BPS condition
and the system becomes tractable under this condition. 
As an example of a system with the BPS condition,
 we can take the Randall-Sundrum model
where the negative cosmological constant
$-6/(\kappa^2\ell^2)$ and brane tension $\sigma$ has the relation 
$\kappa^2\sigma=6/\ell$. 
Assuming this relation, and using a gradient expansion method 
we developed previously~\cite{KS}, we constructed the effective action for 
the Randall-Sundrum model with a bulk scalar field~\cite{Kanno:2003xy}.

In a supersymmetric extension of the Randall-Sundrum model obtained by
 the dimensional reduction of the Ho\v rava-Witten $M$-theory to 
 5-dimensions~\cite{Horava:1995qa}, the bulk 
potential takes an exponential form with
parameters satisfying the BPS condition. 
In the case of homogeneous cosmology, 
the field equations are derived and discussed 
in~\cite{Kobayashi:2002pw}.
For a more thorough analysis, it would be 
nice if a purely 4-dimensional description of the
braneworld exists. Aiming for this direction, 
the effective action for the Ho\v rava-Witten
cosmology is obtained by using 
the moduli approximation~\cite{Garriga}
 (see ~\cite{Palma:2004fh} for other approaches). 

The moduli approximation assumes the metric in the form of a static
solution, but replaces the Minkowski metric $\eta_{\mu\nu}$
 on the brane with a spacetime dependent metric~$g_{\mu\nu}(x)$.
This leads to the metric in a factorized form.
The position of each brane is described by
a spacetime dependent function $\phi(x)$~\cite{Khoury:2001wf}. 
With this form of the metric, the effective action is derived by
 a simple dimensional reduction. 
 However, this factorized metric is not obtained by 
solving the bulk equations of motion. Hence there is no justification
of using it. We should solve the bulk geometry and substitute the
result into the action to obtain a proper 4-dimensional effective
theory. 

In this paper, we will solve the bulk equations
of motion using a gradient expansion method and derive the 4-dimensional 
effective action for the BPS braneworld with a bulk scalar field. 
In the course of analysis, we clarify the relation 
between the gradient expansion method and the moduli approximation method. 
The implications of our result to string cosmology are also discussed. 

The organization of this paper is as follows.
In sec.~2, we present the starting action for our Ho\v rava-Witten
type model, that is, a BPS braneworld with a bulk scalar field.
We then derive basic equations. 
In sec.~3, we explain our iteration scheme of gradient expansion
to solve the Einstein equations, which corresponds to a low energy
approximation, and the background solution is presented. 
In sec.~4, we derive the 4-dimensional effective action.
In sec.~5, the relation to the moduli approximation is investigated. 
The final section is devoted to conclusion. 
In Appendix A, calculation of the extrinsic curvature is given. 

\section{Model And Basic Equations}
We consider a $Z_2$ symmetric 5-dimensional spacetime with two 
branes at the fixed points of the symmetry,
and a bulk scalar field $\varphi$ 
coupled to the brane tension $\sigma(\varphi)$ but not to the matter 
${\cal L}_{\rm matter}$ on the brane. The model is described by the action 
\begin{eqnarray}
S&=&{1\over 2\kappa^2}\int d^5x\sqrt{-\cal G}~
	{\cal R}
	-\int d^5x\sqrt{-\cal G}\left[
	\frac{1}{2}{\cal G}^{AB}\partial_A\varphi\partial_B\varphi
	+U(\varphi)\right]\nonumber\\
&&	-\int d^4x\sqrt{-g_+}\left[
	\sigma(\varphi_+)-{\cal L}^+_{\rm matter}\right]  
	+\int d^4x\sqrt{-g_-}\left[
	\sigma(\varphi_-)+{\cal L}^-_{\rm matter}\right]
	\nonumber\\
&&	+{2\over\kappa^2}\int d^4x\sqrt{-g_+}{\cal K}_+
  	-{2\over\kappa^2}\int d^4x\sqrt{-g_-}{\cal K}_-\ , 
      	\label{action:5-dim}
\end{eqnarray}
where $\kappa^2$ is the 5-dimensional gravitational coupling 
constant and ${\cal R}$ is the 5-dimensional scalar curvature.
We denoted the induced metric on the positive and negative tension branes 
by $g_{\mu\nu}^+$ and $g_{\mu\nu}^-$, respectively. In the last line, we have 
taken into account the Gibbons-Hawking boundary terms instead of 
introducing delta-function singularities in the curvature.
The factor 2 in the Gibbons-Hawking term comes from the $Z_2$ symmetry of 
this spacetime. ${\cal K}_\pm$ is the trace part of the extrinsic curvature
of each boundary brane.
The potential of the bulk scalar field and brane tensions satisfy 
the BPS condition,
\begin{eqnarray}
U(\varphi)=\frac{1}{8}\sigma'^2(\varphi)-\frac{\kappa^2}{6}\sigma^2(\varphi)\ ,
\label{bps}
\end{eqnarray}
where the prime denotes a derivative with respect to the scalar field $\varphi$.
The effective 4-dimensional cosmological constant on the brane will vanish
under this condition. The brane tension is assumed to have the form
\begin{eqnarray}
\sigma=\frac{6\mu}{\kappa^2}
	{\rm exp}\left({\frac{\kappa}{\sqrt{3}p}\varphi}\right)	\ ,
	\label{exponential}
\end{eqnarray}
where $\mu$ and $p$ are model parameters.
 The Ho\v rava-Witten model corresponds to $p=1/\sqrt{6}$.
 The limit $p\rightarrow\infty$ reduces to the Randall-Sundrum model 
with the bulk curvature radius $1/\mu$. 

We take the metric in the bulk as
\begin{eqnarray}
ds^2=\gamma_{\mu\nu}(x,y)dx^\mu dx^\nu+e^{2\psi(x)}dy^2 \ .
\end{eqnarray}
The positive and negative tension branes are respectively placed at
\begin{eqnarray}
y=\phi_+(x)\ ,\qquad y=\phi_-(x) \ ,
\end{eqnarray}
which are often referred to as the moduli fields. In the bulk scalar field
model, we have two scalar degrees of freedom, that is, the bulk scalar 
field and the inter-brane distance. Here we introduced an extra scalar field
$\psi(x)$ as a coordinate choice which is convenient to solve the bulk 
equations of motion. 
Now we give the basic equations in the bulk.
When solving the bulk equations of motion, it is convenient to introduce
a tensor on the $y = {\rm constant} $ slicing by
$K_{\mu\nu}=-1/2~\partial_y\gamma_{\mu\nu}$.
Decomposing the extrinsic curvature into the traceless part and the trace part
\begin{equation}
e^{-\psi}K^{\mu}{}_{\nu}=\Sigma^{\mu}{}_{\nu}+{1\over 4}\delta^{\mu}{}_{\nu}Q \ , \qquad
Q = e^{-\psi}K    \  ,
\label{K} 
\end{equation}
we obtain the basic equations which hold in the bulk;
\begin{eqnarray}
&&e^{-\psi}\Sigma^\mu{}_{\nu ,y}-Q\Sigma^\mu{}_{\nu} 
	=-\left[
	R^\mu{}_\nu(\gamma)  
        -\nabla^\mu\nabla_\nu\psi-\nabla^\mu\psi\nabla_\nu\psi
        -\kappa^2\nabla^\mu\varphi\nabla_\nu\varphi\right]_{\rm traceless}
        \label{evolution1} \ ,     \\
&&e^{-\psi}Q_{,y}-Q^2=\frac{8}{3}\kappa^2U(\varphi)-R(\gamma)
	+\nabla^\alpha\nabla_\alpha\psi
	+\nabla^\alpha\psi\nabla_\alpha\psi
	+\kappa^2\nabla^\alpha\varphi\nabla_\alpha\varphi 
	\label{evolution2} \ , \\
&&e^{-\psi}Q_{,y}-\frac{1}{4}Q^2
	-\Sigma^\alpha{}_{\beta}\Sigma^\beta{}_{\alpha} 
	=\nabla^\alpha\nabla_\alpha\psi
	+\nabla^\alpha\psi\nabla_\alpha\psi
	+\kappa^2\left[e^{-2\psi}(\partial_y\varphi)^2
	+\frac{2}{3}U(\varphi)\right]
	\label{hamiltonian} \ ,  \\
&&\nabla_\lambda\Sigma_{\mu}{}^{\lambda}  
	-{3\over 4}\nabla_\mu Q 
	= -\kappa^2e^{-\psi}(\partial_y\varphi)\nabla_\mu\varphi
	\label{momentum} \ ,\\
&&e^{-\psi}\partial_y\left[e^{-\psi}\partial_y\varphi\right]
	-Qe^{-\psi}\partial_y\varphi
	+\nabla_\alpha\psi\nabla^\alpha\varphi
	+\nabla^{\alpha}\nabla_{\alpha}\varphi-U'(\varphi)=0\ ,
	\label{scalar}
\end{eqnarray}
where $\nabla_\mu $ denotes the covariant differentiation with respect to 
the metric $\gamma_{\mu\nu}$ and $R^\mu{}_\nu(\gamma)$ is the corresponding 
4-dimensional curvature. 
The subscript ``traceless" represents the traceless part of the
quantity in the square brackets.

\section{Gradient Expansion Method}

The effective action has to be derived by substituting the solution of
 Eqs.~(\ref{evolution1})$\sim$(\ref{scalar}) 
into the action (\ref{action:5-dim}) and 
integrating out the result over the bulk coordinate $y$. 
In reality, it is difficult to perform this general procedure.
However, what we need is a low energy effective theory. 
At low energy, the energy density of the matter, $\rho$, 
on the brane is smaller than the brane tension, i.e.,
$\rho /|\sigma| \ll 1$. If we denote the characteristic length scale
of the bulk as $\ell$ and on the brane as $L$,
in this regime, a simple dimensional analysis,
$\rho /|\sigma| \sim \ell^2/L^2\ll 1$, implies that the 4-dimensional curvature
can be neglected compared with the extrinsic curvature. 
Thus, the Anti-Newtonian or gradient expansion method used in the cosmological 
context~\cite{tomita} is applicable to our problem. 
The iteration scheme is to write the metric $\gamma_{\mu\nu}$ 
as a sum of local tensors built out of $g_{\mu\nu}$, 
with the number of derivatives increasing with the order
of iteration, that is, 
$ O((\ell/L)^{2n})$, $n=0,1,2,\cdots$. Hence, we express the metric 
as a perturbative series 
\begin{eqnarray}
&&\gamma_{\mu\nu} (y,x) =
	a^2(y)\left[ g_{\mu\nu} (x) 
  	+f_{\mu\nu}(y,x)
  	+\overset{(2)}{g}_{\mu\nu}
      	+ \cdots  \right]  \ .
      	\label{expansion:metric}
\end{eqnarray}
Other quantities can be also expanded as
\begin{eqnarray}
Q^\mu{}_{\nu}&=&
	\overset{(0)}{Q}{}^{\mu}{}_{\nu}
        +\overset{(1)}{Q}{}^{\mu}{}_{\nu}
	+\overset{(2)}{Q}{}^{\mu}{}_{\nu}+\cdots  \ , \nonumber\\
\Sigma^\mu{}_{\nu}
	&=&\overset{(0)}{\Sigma}{}^{\mu}{}_{\nu}
	+\overset{(1)}{\Sigma}{}^{\mu}{}_{\nu}
	+\overset{(2)}{\Sigma}{}^{\mu}{}_{\nu} + \cdots  \ .
\end{eqnarray}
The bulk equations of motion are to be solved order by
 order~\cite{KS,wiseman}.
\subsection{Zeroth Order}

At zeroth order,  we can neglect the curvature terms in 
Eqs.~(\ref{evolution1})$\sim$(\ref{scalar}). 
Moreover, the tension term only induces the isotropic bending of the brane.
Thus, an anisotropic term vanishes at this order, $
\overset{(0)}{\Sigma}{}^\mu{}_\nu=0$. For simplicity, we write the trace part of the extrinsic curvature
at zeroth order as
\begin{eqnarray}
\overset{(0)}{Q}=W(\overset{(0)}{\varphi}) \ ,
\end{eqnarray}
where $W$ is a function of bulk scalar field at zeroth order.
Combining Eqs. (\ref{evolution2}) with (\ref{hamiltonian}) at this order 
gives
\begin{eqnarray}
&&\frac{3}{4}W^2=\kappa^2e^{-2\psi}\left(
	\partial_y\overset{(0)}{\varphi}\right)^2
	-2\kappa^2U(\overset{(0)}{\varphi}) \ ,
	\label{eq1}\\
&&W'=\frac{4}{3}\kappa^2e^{-\psi}\partial_y\overset{(0)}{\varphi} \ .
	\label{eq2}
\end{eqnarray}
Eliminating $\partial_y\varphi$ from Eq.~(\ref{eq1}) and Eq.~(\ref{eq2}),
we have 
\begin{eqnarray}
W^2=\frac{3}{4\kappa^2}W'^2-\frac{8}{3}\kappa^2U(\varphi) \ .
\label{eqw}
\end{eqnarray}
The remaining Eqs. (\ref{momentum}) and (\ref{scalar}) are automatically
 satisfied.
Comparing Eq.~(\ref{eqw}) with Eqs.~(\ref{bps}) and (\ref{exponential}), 
we get the trace part of the extrinsic curvature in the form
\begin{eqnarray}
W=4\mu{\rm exp}
\left(\frac{\kappa}{\sqrt{3}p}\overset{(0)}{\varphi}\right)
\ .
\end{eqnarray}
As we have obtained the trace part and traceless part of the extrinsic
 curvature,
the extrinsic curvature at this order is given by using Eq.~(\ref{K})
\begin{eqnarray}
e^{-\psi}\overset{(0)}{K}{}^\mu{}_{\nu}
	=\mu\,{\rm exp}\left(
	\frac{\kappa}{\sqrt{3}p}\overset{(0)}{\varphi}\right)
	\delta^{\mu}{}_{\nu} \ .
\label{w}
\end{eqnarray}
Substituting Eq.~(\ref{w}) into Eq.~(\ref{eq2}) and integrating 
with respect to $y$, we obtain the zeroth order scalar field 
\begin{eqnarray}
\overset{(0)}{\varphi}=-\frac{\sqrt{3}p}{\kappa}\log
	\left(1-\frac{\mu}{p^2}y\right)+\eta(x) \ ,
	\label{0:scalar}
\end{eqnarray}
where $\eta(x)$ is a constant of integration and we choose it as
\begin{eqnarray}
\eta(x)=-\frac{\sqrt{3}\,p}{\kappa}\,\psi(x)\ ,
\end{eqnarray}
to make the solution separable. 
Using Eqs.~(\ref{w}), (\ref{0:scalar}) and the definition
\begin{equation}
\overset{(0)}{K}{}_{\mu\nu} 
     =-{1\over 2}\partial_y\overset{(0)}{\gamma}{}_{\mu\nu}     \   ,
\end{equation}
we get the zeroth order metric as
\begin{equation}
ds^2 = a^2 (y) g_{\mu\nu}(x) dx^\mu dx^\nu+e^{2\psi(x)}dy^2 \ ,
\qquad
a(y)=\left(1-\frac{\mu}{p^2}y\right)^{p^2}
\label{0:metric}    \ ,
\end{equation}
where  the tensor $g_{\mu\nu}$ is a constant of integration
 which weakly depends on the brane coordinates $x^\mu$.

\subsection{First Order}

The first order solutions are obtained by taking into account the 
terms neglected at zeroth order. 
At  first order,  Eqs.~(\ref{evolution1})$\sim$(\ref{scalar}) become
\begin{eqnarray}
&&e^{-\psi}\overset{(1)}{\Sigma}{}^{\mu}{}_{\nu , y} 
	-W\overset{(1)}{\Sigma}{}^{\mu}{}_{\nu} 
	=-\left[R^\mu{}_\nu(\gamma) 
	-\nabla^\mu\nabla_\nu\psi-\nabla^\mu\psi\nabla_\nu\psi
        -\kappa^2\nabla^\mu\varphi\nabla_\nu\varphi
        \right]^{(1)}_{\rm traceless}
	\label{1:evolution1} \ , \\
&&e^{-\psi}\overset{(1)}{Q}_{,y}-2W\overset{(1)}{Q}
	=\frac{8}{3}\kappa^2U'(\overset{(0)}{\varphi})\overset{(1)}{\varphi}
	-\left[R(\gamma)-\nabla^\alpha\nabla_\alpha\psi
	-\nabla^\alpha\psi\nabla_\alpha\psi
	-\kappa^2\nabla^\alpha\varphi\nabla_\alpha\varphi\right]^{(1)} 
	\label{1:evolution2} \ , \\
&&e^{-\psi}\overset{(1)}{Q}_{,y}-\frac{1}{2}W\overset{(1)}{Q}
	=\left[\nabla^\alpha\nabla_\alpha\psi
	+\nabla^\alpha\psi\nabla_\alpha\psi\right]^{(1)}
	+\kappa^2\left[2e^{-2\psi}
	(\partial_y\overset{(0)}{\varphi})(\partial_y\overset{(1)}{\varphi})
	+\frac{2}{3}U'(\overset{(0)}{\varphi})\overset{(1)}{\varphi}\right]
	\label{1:hamiltonian} \ ,  \\
&&\nabla_\lambda\overset{(1)}{\Sigma}{}_{\mu}{}^{\lambda}{}  
	-{3\over 4}\nabla_\mu\overset{(1)}{Q}=
	-\kappa^2e^{-\psi}\left(
	\partial_y\overset{(0)}\varphi\partial_\mu\overset{(1)}\varphi
	+\partial_y\overset{(1)}\varphi\partial_\mu\overset{(0)}\varphi
	\right) 
	\label{1:momentum}\ ,\\
&&e^{-\psi}\partial_y\left[e^{-\psi}\partial_y\overset{(1)}{\varphi}\right]
	-\overset{(1)}{Q}e^{-\psi}\partial_y\overset{(0)}{\varphi}
	-We^{-\psi}\partial_y\overset{(1)}{\varphi}
	+\nabla^\alpha\nabla_\alpha\overset{(0)}{\varphi}
	+\partial_\mu\psi\partial^\mu\overset{(0)}{\varphi}
	-U''(\overset{(0)}{\varphi})\overset{(1)}{\varphi}=0 \ ,
	\label{1:scalar}
\end{eqnarray}
where the superscript $(1)$ represents the order of the derivative expansion.
Here, $[R^\mu{}_\nu(\gamma)]^{(1)} $ 
means that the curvature is approximated by
 taking the Ricci tensor of $a^2(y)g_{\mu\nu}(x)$ 
 in place of the full metric $\gamma_{\mu\nu}(x,y)$.
 It is also convenient to write it in terms of the Ricci
  tensor of $g_{\mu\nu}$, denoted by $R^\mu{}_\nu (g)$.
 
Combining Eq.~(\ref{1:evolution2}) with Eq.~(\ref{1:hamiltonian}), 
we get 
\begin{eqnarray}
\overset{(1)}{Q}=\frac{4\kappa^2}{3W}\left[e^{-2\psi}
	(\partial_y\overset{(0)}{\varphi})(\partial_y\overset{(1)}{\varphi})
	-U'(\overset{(0)}{\varphi})\overset{(1)}{\varphi}\right]
	+\frac{2}{3W}\left[R(\gamma)-\kappa^2
	\nabla^\alpha\overset{(0)}{\varphi}
	\nabla_\alpha\overset{(0)}{\varphi}\right]^{(1)}\ .
	\label{1:Q}
\end{eqnarray}
Using Eq.~(\ref{eq2}) and substituting Eq.~({\ref{1:Q}}) into 
Eq.~(\ref{1:scalar}), we obtain
\begin{eqnarray}
&&e^{-2\psi}\partial_y^2\overset{(1)}{\varphi}
	-\left(\frac{3}{4\kappa^2}\frac{W'^2}{W}+W
	\right)e^{-\psi}\partial_y\overset{(1)}{\varphi}
	+\left(\frac{W'}{W}U'-U''
	\right)\overset{(1)}{\varphi}\nonumber\\
&&\qquad\qquad
	+\nabla^\alpha\nabla_\alpha\overset{(0)}{\varphi}
	+\partial_\alpha\psi\partial^\alpha\overset{(0)}{\varphi}
	-\frac{1}{2\kappa^2}\frac{W'}{W}
	\left[R(\gamma)-\kappa^2
	\nabla^\alpha\overset{(0)}{\varphi}
	\nabla_\alpha\overset{(0)}{\varphi}\right]^{(1)}
	=0 \ ,
\label{scalar1st}
\end{eqnarray}
where the potential $U(\varphi)$ appearing in front of the 
first order quantity is given by Eq.~(\ref{bps})
\begin{eqnarray}
U(\varphi)=-\frac{6\mu^2}{\kappa^2}\left(1-\frac{1}{4p^2}\right)
	{\rm exp}\left(\frac{2\kappa}{\sqrt{3}p}\overset{(0)}{\varphi}\right)
	\ .
\end{eqnarray}
The last term is approximated by substituting the zeroth order metric 
(\ref{0:metric}),
\begin{eqnarray}
\left[R(\gamma)-\kappa^2
	\nabla^\alpha\overset{(0)}{\varphi}
	\nabla_\alpha\overset{(0)}{\varphi}\right]^{(1)}
	=\frac{1}{a^2(y)}\left[R(g)-3p^2\psi^{|\alpha}\psi_{|\alpha}
	\right] \ ,
\end{eqnarray}
where $|$ denotes the covariant differentiation with respect to
the metric $g_{\mu\nu}$. Hereafter we omit the argument of the curvature 
and describe $\psi^{|\alpha}\psi_{|\alpha}$ as $(\partial\psi)^2$ for 
simplicity.

Substituting the zeroth order solutions (\ref{w}), (\ref{0:scalar})
and (\ref{0:metric}) into Eq.~(\ref{scalar1st}), 
we obtain the scalar field at this order
\begin{eqnarray}
\overset{(1)}{\varphi}&=&\beta(x) a^{-2+2/p^2}(y)
	+C_1(x)a^{-1/p^2}(y)+C_2(x)a^{-4+1/p^2}(y)\ ,
\end{eqnarray}
where $C_1$ and $C_2$ are constants of integration 
which are functions of $x$.
The function $\beta$ is a particular solution given by
\begin{eqnarray}
\beta(x)=-\frac{p^4}{16\mu^4(2p^2+1)(2p^2-3)}e^{2\psi}
	\left[\frac{\sqrt{3}p}{\kappa}\square\psi
	+\frac{1}{2\sqrt{3}p\kappa}
	\left(R+3p^2(\partial\psi)^2
	\right)\right]\ ,
	\label{beta}
\end{eqnarray}
where the d'Alembertian $\square$ is with respect to the metric $g_{\mu\nu}$.
We note that $C_1$ and $C_2$ are determined by the junction conditions at
the two boundary branes. It may be noted that because
we have introduced an auxiliary field $\psi$ in the metric,
we may choose to set $\overset{(1)}{\varphi}=0$ at the
positive tension brane, if desired.
Substituting this solution into Eq.~(\ref{1:Q}), we get
\begin{eqnarray}
\overset{(1)}{Q}&=&
	\frac{1}{6\mu}e^{\psi}a^{1/p^2-2}
	\left[R-3p^2(\partial\psi)^2\right]
	+\frac{\kappa \mu}{\sqrt{3}p^3}
	e^{-\psi}\left[
	(6p^2-3)a^{1/p^2-2}\beta 
	+4p^2a^{-2/p^2}C_1
	+(8p^2-2)a^{-4}C_2\right]
	\label{1:trc} \ .
\end{eqnarray}
Using Eq.~(\ref{1:evolution1}), we obtain
\begin{eqnarray}
\overset{(1)}{\Sigma}{}^{\mu}{}_{\nu}=\frac{p^2}{\mu(2p^2+1)}
	e^{\psi}a^{1/p^2-2}
	\left[R^\mu{}_\nu(g)-\Box\psi
	-(3p^2+1)\psi^{|\mu}\psi_{|\nu}
	\right]_{\rm traceless}
	+a^{-4}\chi^\mu{}_\nu(x)\ ,
	\label{1:trclss}
\end{eqnarray}
where $\chi^\mu{}_\nu$ is a constant of integration which satisfies 
\begin{eqnarray}
\chi^{\mu}{}_{\mu}=0   \ , \quad\chi^{\mu}{}_{\nu|\mu}=0 \ .
\label{TT}
\end{eqnarray}
Here, the latter condition came from Eq.~(\ref{1:momentum}). 
We note that $\chi^\mu{}_\nu$ is by itself a physical quantity,
and determined by the junction condition~\cite{KS,Kanno:2003xy}.
In fact, this term corresponds to the dark radiation at this order~\cite{KS}.
We also note that, unlike the other quantities, due to the traceless property,
$\chi^\mu{}_{\nu}$ does not appear in the derivation of the 
effective action at first order.

The definition of the extrinsic curvature at this order is expressed as
\begin{eqnarray}
e^{-\psi}\overset{(1)}{K}{}^{\mu}{}_{\nu}=
	-\frac{1}{2}g^{\mu\alpha}e^{-\psi}
	\partial_yf_{\alpha\nu}\ .
\end{eqnarray}
{}From Eqs.(\ref{1:trc}) and  (\ref{1:trclss}),
the correction to the metric $g_{\mu\nu}$ at this order can be obtained as
\begin{eqnarray}
f_{\mu\nu}(y,x)&=&-\frac{p^4}{\mu^2(p^2-1)(2p^2+1)}e^{-2\psi}a^{-2+2/p^2}
	\left[ R_{\mu\nu}-\psi_{|\mu\nu}
	-(3p^2+1)\psi_{|\mu}\psi_{|\nu}\right]_{\rm traceless}
	\nonumber\\
&&	-\frac{p^2}{24\mu^2(p^2-1)}a^{2\psi}g_{\mu\nu}\left(
	R-3p^2(\partial\psi)^2\right)
	-\frac{2p^2}{\mu(4p^2-1)}e^{\psi}a^{-4+1/p^2}\chi_{\mu\nu}
	+C_{\mu\nu}(x)\nonumber\\
&&	-\frac{\kappa}{2\sqrt{3}}g_{\mu\nu}\left(
	\frac{3(2p^2-1)}{2p(p^2-1)}a^{-2+2/p^2}\beta
	+3pa^{-1/p^2}C_1
	+\frac{2}{p}a^{-4+1/p^2}C_2\right) 
	\label{1:metric}\ ,
\end{eqnarray}
where $C_{\mu\nu}$ is another constant of integration. 
Together with $\chi_{\mu\nu}$, it is also determined by the 
junction conditions. However, since there is freedom to
specify the metric form at either of the branes, we may
choose it to set the induced metric on, say, the positive
tension brane as
\begin{eqnarray}
g^{+}_{\mu\nu}(x)=\gamma_{\mu\nu}\left(\phi_+(x),x\right)=a_+^2\,g_{\mu\nu}(x).
\end{eqnarray}

Substituting the metric up to first order,
as given by Eq.~(\ref{expansion:metric}), into
the original action~(\ref{action:5-dim}), we get the 4-dimensional 
effective action at leading order. The Kaluza-Klein (KK) corrections
start from the 2nd order in our method. In principle, it is possible
to solve the bulk equations of motion 
up to second order~\cite{KS,Kanno:2003vf}. 
However, it is beyond the scope of this paper.
In the next section, we will focus on constructing the effective action 
at leading order.

\section{Effective Action}
The effective action can be constructed with the knowledge of
the leading order metric $f_{\mu\nu}(y,x)$. 
Now, up to the first order, we have
\begin{eqnarray}
\gamma_{\mu\nu}(y,x)=a^2(y)
	\left[
	g_{\mu\nu}(x)+f_{\mu\nu}(y,x)
	\right]    \ . 
\end{eqnarray}
In the following, we will calculate the bulk action, 
$S_{\rm bulk}$, 
the actions for each brane tensions, $S_\pm$ , 
brane matter $S^\pm_{\rm m}$ and the 
Gibbons-Hawking term, $S_{\rm GH}$, separately. After that, we collect all of 
them and obtain the 4-dimensional effective action.

In order to calculate the bulk action, 
we need the determinant of the bulk metric 
\begin{eqnarray}
\sqrt{\cal -G}&=&e^{\psi}a^4(y)\sqrt{-g}\sqrt{1+{\rm tr}f}
	\nonumber\\
&\approx&
	e^{\psi}a^4(y)\sqrt{-g}\left(
	1+\frac{1}{2}{\rm tr}f\right) \ , 
\end{eqnarray}
where
\begin{eqnarray}
{\rm tr} f&=&-\frac{p^2}{6\mu^2(p^2-1)}e^{2\psi}a^{-2+2/p^2}\left(
	R-3p^2(\partial\psi)^2\right)\nonumber\\
&&	-\frac{\sqrt{3}\kappa(2p^2-1)}{p~(p^2-1)}a^{-2+2/p^2}\beta
	-\frac{8\kappa}{\sqrt{3}}p\,a^{-1/p^2}C_1
	-\frac{4\kappa}{\sqrt{3}\,p}a^{-4+1/p^2}C_2+C^\mu{}_\mu\ .
\end{eqnarray}
As we have solved the bulk equations of motion, we can use the 
the equation 
${\cal R}=\kappa^2({\cal G}^{AB}\partial_A\varphi\partial_B\varphi
+10/3U(\varphi)) $ which holds in the bulk. 
Then, the bulk action becomes
\begin{eqnarray}
S_{\rm bulk}&\equiv&{1\over 2\kappa^2}\int d^5x\sqrt{\cal -G}~
	{\cal R}
	-\int d^5x\sqrt{-\cal G}\left[
	\frac{1}{2}{\cal G}^{AB}\partial_A\varphi\partial_B\varphi
	+U(\varphi)\right]
	\nonumber\\
&=&-\frac{2}{\kappa^2}\int d^4x\sqrt{-g}
	\left[
	\left\{\frac{p^2(4p^2-1)}{12\mu(p^2-1)(2p^2+1)}e^{\psi}
	\left(R-3p^2(\partial\psi)^2\right)
+\frac{\kappa \mu(4p^2-1)}
	{2\sqrt{3}p(p^2-1)}\,e^{-\psi}\beta\right\}a^{2+1/p^2}(y)
\right.
\nonumber\\
&&
	\left.\hspace{2.5cm}
-\mu\,e^{\psi}\,a^{4-1/p^2}(y)
	+\frac{\kappa \mu(4p^2-1)}{\sqrt{3}p}\,
	e^{-\psi}C_1\,a^{4-2/p^2}(y)
	\right]^{\phi_-}_{\phi_+}\,\left(1+\frac{C^\mu{}_\mu}{2}\right) \ ,
	\label{bulk}
\end{eqnarray}
where the factor 2 over $\kappa^2$ comes from the $Z_2$ symmetry of 
this spacetime and 
we neglected the second order quantities. 
Notice that the Ricci scalar came from ${\rm tr}f_{\mu\nu}$ 
in $\sqrt{\cal -G}$.

Next, let us calculate the action for the brane tension. 
 The induced metric on each 
brane is  written by
\begin{eqnarray}
g^\pm_{\mu\nu}(\phi_\pm,x)=
	a_{\pm}^2g_{\mu\nu}(x)+a_{\pm}^2f_{\mu\nu}(\phi_\pm,x)
	+\partial_\mu\phi_\pm\partial_\nu\phi_\pm  \ ,
\end{eqnarray}
where $a_{\pm}=a(\phi_{\pm})$.
The determinant of the induced metric can be calculated as 
\begin{eqnarray}
\sqrt{-g_\pm}&=&a^4_{\pm}\sqrt{-g}
	\sqrt{1+\frac{1}{a_{\pm}^2}e^{2\psi}(\partial\phi_\pm)^2
	+{\rm tr}f}
	\nonumber\\
&\approx&
	a_{\pm}^4\sqrt{-g}\left(
	1+\frac{1}{2a_{\pm}^2}e^{2\psi}(\partial\phi_\pm)^2
	+\frac{1}{2}{\rm tr} f\right)
	\label{induced} \ ,
\end{eqnarray}
The brane tension given by Eq.~(\ref{bps}) is approximated
up to first order as
\begin{eqnarray}
\sigma(\varphi)\approx
	e^{-\psi}a^{-1/p^2}
	+\frac{\kappa}{\sqrt{3}p}e^{-\psi}\left(a^{-2+1/p^2}\beta
	+a^{-2/p^2}C_1+a^{-4}C_2\right)\ .
\end{eqnarray}
Thus, the action for each brane becomes
\begin{eqnarray}
S_\pm &\equiv& \mp \int d^4x\sqrt{-g_\pm}~\sigma(\varphi)  
	\nonumber\\
&=&\mp\frac{1}{\kappa^2}\int d^4x\sqrt{-g}
	\left[
	-\frac{p^2}{2\mu(p^2-1)}e^{\psi}a_{\pm}^{2+1/p^2}
	\left(R-3p^2(\partial\psi)^2\right)
	+6\mu\,e^{-\psi}a_{\pm}^{4-1/p^2}
	+3\mu\,e^{\psi}a_{\pm}^{2-1/p^2}(\partial\phi_\pm )^2
	\right.\nonumber\\
&&	\left.
	-\frac{\sqrt{3}\kappa \mu(4p^2-1)}{p(p^2-1)}e^{-\psi}
	a_\pm^{2+1/p^2}\beta
	-\frac{2\sqrt{3}\kappa \mu(4p^2-1)}{p}e^{-\psi}
	a_\pm^{4-2/p^2}C_1
	+\frac{2\sqrt{3}\kappa \mu}{p}e^{-\psi}~C_2
	~\right]\left[
	1+\frac{C^\mu{}_\mu}{2}\right]  \ .
	\label{tension}
\end{eqnarray}
Note that the Ricci scalar came from ${\rm tr}f_{\mu\nu}$
in $\sqrt{-g_\pm}$.

As to the brane matter which corresponds to a first order quantity,
we take the zeroth order metric in Eq.~(\ref{induced})
\begin{eqnarray}
S_{\rm m}^{\pm} &\equiv&\int d^4x\sqrt{-g_\pm}~{\cal L}_{\rm matter}
  	\nonumber\\
&=&\int d^4x\sqrt{-g}~a_\pm^4{\cal L}_{\rm matter} \ .
\label{matter}
\end{eqnarray}

In order to calculate the Gibbons-Hawking term, we need the extrinsic
curvature on the $y=\phi(x)$ slicing defined by
\begin{eqnarray}
{\cal K}_{\mu\nu}\equiv n_A\left(
	\frac{\partial^2x^A}{\partial\xi^\mu\partial\xi^\nu}
	\right)
	+\Gamma^A_{BD}
	\frac{\partial x^B}{\partial\xi^\mu}
	\frac{\partial x^D}{\partial\xi^\nu}  \ ,
\end{eqnarray}  
where $x^A$ is the coordinate of the brane, $\xi^\mu = x^\mu $ 
is the one on the brane 
and $n_A$ is the normal vector to the brane. Note that ${\cal K}_{\mu\nu}$ 
is different from $K_{\mu\nu}$ in Eq.~(\ref{K}). For the detail calculation
see Appendix A. The trace part of the extrinsic curvature on each brane is 
obtained as
\begin{eqnarray}
{\cal K}_\pm&=&g^{\mu\nu}_\pm{\cal K}^\pm_{\mu\nu}\nonumber\\
	&=&n_y\left[
	4\mu\,e^{-2\psi}a_\pm^{-1/p^2}+a_\pm^{-2}\square\phi_\pm
	+\mu\,a_\pm^{-2-1/p^2}(\partial\phi_\pm)^2
	+2a_\pm^{-2}(\partial_\alpha\phi)(\partial^\alpha\psi)
	+\frac{1}{6\mu}a_\pm^{-2+1/p^2}\left(R-3p^2(\partial\psi)^2\right)
	\right.\nonumber\\
&&	\left.\qquad
	+\frac{\kappa}{\sqrt{3}}e^{-2\psi}\left\{
	\frac{3(2p^2-1)\mu}{p^3}a_\pm^{-2+1/p^2}\beta
	+\frac{4\mu}{p}a_\pm^{-2/p^2}C_1
	+\frac{2(4p^2-1)\mu}{p^3}a_\pm^{-4}C_2
	\right\}\right] \ .
\end{eqnarray}
Thus, the Gibbons-Hawking term is obtained as
\begin{eqnarray}
S_{\rm GH}&\equiv&{2\over\kappa^2}\int d^4x\sqrt{-g_+}{\cal K}_+
	-{2\over\kappa^2}\int d^4x\sqrt{-g_-}{\cal K}_-
	\nonumber\\
&=&	{2\over\kappa^2}\int d^4x\sqrt{-g}\left[
	-\frac{(p^2+1)}{6\mu(p^2-1)}e^{\psi}a_+^{2+1/p^2}
	\left(R-3p^2(\partial\psi)^2\right)
	+3\mu\,e^{\psi}a_+^{2-1/p^2}(\partial\phi_+)^2
	\right.\nonumber\\
&&	\left.\hspace{2cm}
	+e^{\psi}a_+^{2}(\partial_\alpha\phi)(\partial^\alpha\psi)
	+4\mu\,e^{-\psi}a_+^{4-1/p^2}
	-\frac{\sqrt{3}\kappa \mu(p^2+1)(2p^2-1)}{p^3(p^2-1)}
	e^{-\psi}a_+^{2+1/p^2}\beta
	\right.\nonumber\\
&&	\left.\hspace{2cm}
	-\frac{4\kappa \mu(4p^2-1)}{\sqrt{3}p}e^{-\psi}a_+^{4-2/p^2}C_1
	-\frac{2\kappa \mu}{\sqrt{3}p^3}e^{-\psi}~C_2	
	~\right]\left[
	1+\frac{C^\mu{}_\mu}{2}\right]
	+\left(\phi_+\rightarrow\phi_-\right) \ .
	\label{GH}
\end{eqnarray}
Note that the Ricci scalar came from ${\rm tr}f_{\mu\nu}$ in
$\sqrt{-g_\pm}$ and ${\rm tr}f_{\mu\nu,y}$ in ${\cal K}_\pm$.

Substituting Eqs.~(\ref{bulk}), (\ref{tension}),
(\ref{matter}) and (\ref{GH})
into the 5-dimensional action Eq.~(\ref{action:5-dim}), 
we get the 4-dimensional effective action
\begin{eqnarray}
S&=&S_{{\rm bulk}+\varphi}+S_++S_-+S^+_{\rm m}+S^-_{\rm m}+S_{\rm GH}\nonumber\\&=&	\frac{1}{2\kappa^2\mu}\int d^4x\sqrt{-g}e^{\psi}
	\left[
	(\Upsilon_+^2-\Upsilon_-^2)R+\frac{12p^2}{2p^2+1}\left\{
	(\partial\Upsilon_+)^2-(\partial\Upsilon_-)^2\right\}
	-3p^2(\Upsilon_+^2-\Upsilon_-^2)(\partial\psi)^2
	\right]\left[
	1+\frac{C^\mu{}_\mu}{2}\right]
	\nonumber\\
&&	+\left(
	\frac{(2p^2+1)}{2p^2}\right)^{\frac{4p^2}{(2p^2+1)}}
	\int d^4x\sqrt{-g}~\left[\Upsilon_+^{\frac{8p^2}{(2p^2+1)}}
	{\cal L}^+_{\rm matter}-
	\Upsilon_-^{\frac{8p^2}{(2p^2+1)}}{\cal L}^-_{\rm matter}
	\right]   
	\label{4d}\ .
\end{eqnarray}
Here we have used Eq.~(\ref{beta}) and introduced
\begin{eqnarray}
\Upsilon_+^2=\frac{2p^2}{(2p^2+1)}a_+^{2+1/p^2}\,,\qquad
\Upsilon_-^2=\frac{2p^2}{(2p^2+1)}a_-^{2+1/p^2}\,.
\label{upsilon1}
\end{eqnarray}
Note that $C^\mu{}_\mu$ is a first order quantity, so we can ignore
this term at leading order.
It should be stressed that the Einstein-Hilbert term is originated from
the contributions of $f_{\mu\nu}$ in each 
$S_{\rm bulk}$, $S_\pm$ and $S_{\rm GH}$,
 so the correction $f_{\mu\nu}$ to the 
 metric $g_{\mu\nu}$ plays an important role. 
The kinetic term of the moduli field arising from the 
position of the positive tension bran
seems to have a wrong sign.
However, this may be simply because the action is not in the 
Einstein-Hilbert form.
To see if it is a ghost or not, we have to go to the 
Einstein frame by performing a conformal transformation.
 It is convenient
to introduce the new moduli fields, $\Psi$ and $\Phi$,
defined by
\begin{eqnarray}
\Upsilon_+=e^{-(p^2+1/2)\psi}\Psi{\rm cosh}\Phi \ ,\qquad 
\Upsilon_-=e^{-(p^2+1/2)\psi}\Psi{\rm sinh}\Phi \ .
\label{upsilon2}
\end{eqnarray}
and introduce the metric in the Einstein frame,
\begin{eqnarray}
{\tilde g}_{\mu\nu}=e^{-2p^2\psi}\Psi^2g_{\mu\nu} \ .
\end{eqnarray}
Then, we find
\begin{eqnarray}
S&=&\frac{1}{2\kappa^2\mu}\int d^4x\sqrt{-\tilde{g}}
	\left[R(\tilde{g}) 
	-\frac{6}{2p^2+1}\frac{(\partial\Psi)^2}{\Psi^2}
	-\frac{12p^2}{2p^2+1}(\partial\Phi)^2
	\right]\nonumber\\
&&	+\left(\frac{(2p^2+1)}
	{2p^2}\right)^{\frac{4p^2}{(2p^2+1)}}
	\int d^4x\sqrt{-\tilde{g}}\Psi^{-\frac{4}{2p^2+1}}
	\left[
	\left({\rm cosh\Phi}\right)^{\frac{8p^2}{(2p^2+1)}}
	{\cal L}^+_{\rm matter}
	-\left({\rm sinh\Phi}\right)^{\frac{8p^2}{(2p^2+1)}}
	{\cal L}^-_{\rm matter}\right] \ .
	\label{einstein}
\end{eqnarray}
Therefore, both moduli have positive kinetic terms in the Einstein
frame. 
It should be noted that in going to the Einstein frame, 
the metric $\tilde{g}_{\mu\nu}$ is written by the induced metric 
with a conformal factor,
\begin{eqnarray}
\tilde{g}_{\mu\nu}&=&\left(
	\frac{(2p^2+1)}{2p^2}\right)^{-2p^2/(2p^2+1)}
	\Psi^{2/(2p^2+1)}
	\left({\rm cosh\Phi}\right)^{-4p^2/(2p^2+1)}
	g_{\mu\nu}^+\nonumber \ ,\\
\tilde{g}_{\mu\nu}&=&\left(
	\frac{(2p^2+1)}{2p^2}\right)^{-2p^2/(2p^2+1)}
	\Psi^{2/(2p^2+1)}
	\left({\rm sinh\Phi}\right)^{-4p^2/(2p^2+1)}
	g_{\mu\nu}^- \ .
\label{EtoInd}
\end{eqnarray}
Then, in the Einstein frame the action for matter has the form
\begin{eqnarray}
{\cal L}^+_{\rm matter}={\cal L}^+_{\rm matter}(\tau_i,g_+
(\Psi,\Phi,\tilde{g}_{\mu\nu}))\ ,\qquad
{\cal L}^-_{\rm matter}={\cal L}^-_{\rm matter}(\tau_i,g_-
(\Psi,\Phi,\tilde{g}_{\mu\nu}))
\end{eqnarray}
where $\tau_i$ denotes the matter fields on each brane which do not
couple to the bulk scalar field.

\section{Comparison with Moduli Approximation}
In our previous paper~\cite{Kanno:2004yb}, we investigated a toy 
problem in the Randall-Sundrum two-brane model to confirm the validity of the 
factorizable metric ansatz used in the moduli approximation. 
Here we also investigate the validity of the factorizable metric ansatz 
in more realistic models from the stringy point of view. 
As we see from Eqs.~(\ref{0:metric}) and 
(\ref{1:metric}), the metric to first order looks
quite different from the factorizable form. 
Now let us recapitulate the derivation
of the 4-dimensional effective
action used in moduli approximation. 

In the moduli approximation, one assumes the following factorizable metric
which is obtained by replacing the 4-dimensional Minkowski metric on the brane
of the static solution with a spacetime dependent metric $g_{\mu\nu}$,
\begin{eqnarray}
  ds^2 = a^2 (y) g_{\mu\nu} (x) dx^\mu dx^\nu + dy^2
  \label{factorized} \ .
\end{eqnarray}
The positions of the branes are assumed to fluctuate as 
\begin{eqnarray}
y=\phi_+(x),\qquad y= \phi_-(x)  \ ,
\end{eqnarray}
For the metric~(\ref{factorized}), using the Gauss equation,
it is straightforward to relate the bulk Ricci scalar to the
 4-dimensional Ricci scalar,
\begin{eqnarray}
{\cal R}=a^{-2}R(g)+\left(\frac{A^2}{2p^2}-\frac{5A^2}{4}\right)a^{-2/p^2}
\end{eqnarray}
Following the same procedure that leads to Eq.~(\ref{4d}) 
with the factorized metric (\ref{factorized}), 
we get the action
\begin{eqnarray}
S&=&\frac{1}{2\kappa^2\mu}\int d^4x\sqrt{-g}
	\left[\left(
	\Upsilon_+^2-\Upsilon_-^2\right)R(g)
	+\frac{12p^2}{2p^2+1}\left\{
	(\partial\Upsilon_+)^2-(\partial\Upsilon_-)^2\right\}\right]
	\nonumber\\
&&	+\left(
	\frac{(2p^2+1)}{2p^2}\right)^{\frac{4p^2}{(2p^2+1)}}
	\int d^4x\sqrt{-g}~\left[\Upsilon_+^{\frac{8p^2}{(2p^2+1)}}
	{\cal L}^+_{\rm matter}-
	\Upsilon_-^{\frac{8p^2}{(2p^2+1)}}{\cal L}^-_{\rm matter}
	\right]   \ .
	\label{moduliapp}
\end{eqnarray}
where we used the same variable defined in (\ref{upsilon1}).
Apparently, the effective action (\ref{moduliapp}) looks
different from the action~(\ref{4d}).
However we should not jump to conclusion. We should compare
the results in the Einstein frame.
Then using the variables, 
\begin{eqnarray}
\Upsilon_+=\Psi{\rm cosh}\Phi \ ,\qquad 
\Upsilon_-=\Psi{\rm sinh}\Phi \ .
\end{eqnarray}
and performing a conformal transformation
\begin{eqnarray}
{\tilde g}_{\mu\nu}=\Psi^2g_{\mu\nu}
\end{eqnarray}
We have the action in the Einstein frame
\begin{eqnarray}
S&=&\frac{1}{2\kappa^2\mu}\int d^4x\sqrt{-\tilde{g}}
	\left[R(\tilde{g}) 
	-\frac{6}{2p^2+1}\frac{(\partial\Psi)^2}{\Psi^2}
	-\frac{12p^2}{2p^2+1}(\partial\Phi)^2
	\right]\nonumber\\
&&	+\left(\frac{(2p^2+1)}
	{2p^2}\right)^{\frac{4p^2}{(2p^2+1)}}
	\int d^4x\sqrt{-\tilde{g}}\tilde\Psi^{-4/(2p^2+1)}
	\left[
	\left({\rm cosh\Phi}\right)^{\frac{8p^2}{(2p^2+1)}}
	{\cal L}^+_{\rm matter}
	-\left({\rm sinh\Phi}\right)^{\frac{8p^2}{(2p^2+1)}}
	{\cal L}^-_{\rm matter}\right] \ .
	\label{moduliapp-einstein}
\end{eqnarray}
We see that this effective action (\ref{moduliapp-einstein}) is 
indistinguishable from Eq.~(\ref{einstein}) 
obtained by solving the bulk equations of motion.
Thus, we have shown that the action obtained from the moduli
approximation is correct at the leading order.

It may be noted that, because the zeroth order term, that is,
a cosmological constant is absent in the effective action
 as a result of the BPS condition,
the difference between the induced metric on each of the branes
obtained from the factorized metric~(\ref{factorized}) in the moduli
 approximation and the induced metric in our gradient expansion approach,
 which is of first order, turns out to be irrelevant~\cite{Kanno:2004yb}.

Recent intense research on inflationary models in string theory 
has stemmed from the success of constructing a model of 
brane inflation by moduli stabilization~\cite{Giddings:2001yu}.
Almost all studies which compute potentials for moduli 
in type IIB string theory start from the factorizable ansatz for
the 10-dimensional metric. Hence it is important to verify
the factorizable ansatz or the moduli approximation in general.
Our calculation supports the validity of the factorizable metric
 ansatz and the moduli approximation in string theory
at sufficiently low energies.

\section{Conclusion}

As a supersymmetric extension of the Randall-Sundrum
model, we have considered a 5-dimensional Ho\v rava-Witten type theory,
and derived its low energy effective action.
The model consists of two boundary branes and a bulk scalar field
and the potential of the bulk scalar field is related to the 
tension of the branes in such a way that the BPS condition is
satisfied. 

The effective action is obtained by solving the bulk equations of motion
and substituting the result into the original action.
We have used the gradient expansion method~\cite{KS,Kanno:2003xy},
which is valid at low energies, and solved the bulk to first order
 in the expansion.
The resultant theory can be cast into the form of Einstein gravity
coupled with two scalar fields, one arising from the inter-brane distance
degree of freedom, the radion, and the other from the bulk scalar field.
It is known that the Ho\v rava-Witten type theory can be
 derived from the 6-dimensional Randall-Sundrum model~\cite{Kobayashi:2003cb}.
It would be of interest to derive our effective action from the 
6-dimensional point of view.

We have also shown that the 4-dimensional effective action we obtained by
the gradient expansion to first order is equivalent to the one by
the moduli approximation. Hence, our result supports
the moduli approximation and also supports the factorizable
metric ansatz used in string cosmology at sufficiently low
energies.

In the current study of string cosmology, 
only the leading order effect has been discussed.
At this order, as we have shown, the moduli approximation
seems to be valid and useful.
Undoubtedly, however, the next order effect is of
great interest because it is where the genuine bulk
degrees of freedom, the so-called Kaluza-Klein
modes, start to play a role. In such a case,
neither the moduli approximation nor any extension of it
is reliable. One of the reasons is that the factorizable
metric ansatz is no longer valid~\cite{Kanno:2003vf}.
Our approach, based on the gradient expansion, will
be very useful for the evaluation of high energy
corrections.

\begin{acknowledgements}
I am very grateful to Jiro Soda for invaluable discussions, advice and
continuous encouragement. 
I would like to thank Misao Sasaki for useful discussions and suggestions.
 This work was initiated while I was visiting ICG at Portsmouth University. 
I am grateful to ICG for financial support and the ICG members
 for warm hospitality.
I have especially benefitted from discussions with
 Roy Maartens and David Wands.
This work was supported in part by JSPS Grant-in-Aid for Scientific
Research, No.155476 and also by the 21COE program ``Center for
  Diversity and Universality in Physics", Kyoto University.
\end{acknowledgements}

\appendix
\section{Extrinsic Curvature}
The Christoffel symbols 
we need are
\begin{eqnarray}
\Gamma^y_{y\mu}&=&\psi_{,\mu}\ \ ,\qquad
\Gamma^y_{\mu\nu}=\mu\,e^{-2\psi}a^{2-1/p^2}\left(
	g_{\mu\nu}+f_{\mu\nu}\right)
	-\frac{1}{2}e^{-2\psi}a^2f_{\mu\nu,y}\ \ , \\
\Gamma^{\mu}_{yy}&=&-e^{2\psi}\psi^{,\mu}\ \ ,\qquad
\Gamma^\alpha_{y\mu}=-\mu\,a^{-1/p^2}\delta^\alpha_\mu
	+\frac{1}{2}g^{\alpha\beta}f_{\beta\mu,y}\ \ ,\qquad
\Gamma^{\mu}_{\nu\lambda}=\Gamma^{\mu}_{\nu\lambda}(\gamma) \ .
\end{eqnarray}
The tangent basis on the brane are given by
\begin{eqnarray}
\frac{\partial x^A}{\partial\xi^\mu}
=(\delta^\alpha_\mu,~\partial_\mu\phi_\pm) \ .
\end{eqnarray}
Thus, the normal vector takes the form 
\begin{eqnarray}
n_A=(-n_y\partial_\alpha\phi_\pm,~n_y) \ .
\end{eqnarray}
{}From the normalization condition $n_A n^A =1$, we have
\begin{eqnarray}
n_y=e^{\psi}\frac{1}{\sqrt{1
	+a_{\pm}^{-2}e^{2\psi}(\partial\phi_\pm)^2}} \ ,
\end{eqnarray}
where $a_\pm=a(\phi_\pm)$ as defined in the text.
Then the extrinsic curvature is calculated as
\begin{eqnarray}
{\cal K}^\pm_{\mu\nu}&=&n_y\left[
	\nabla_\mu\nabla_\nu\phi_\pm
	+2\partial_\mu\psi\partial_\nu\phi_{\pm}
	+\mu\,e^{-2\psi}a_\pm^{2-1/p^2}\left(
	g_{\mu\nu}+f_{\mu\nu}
	\right)
	-\frac{1}{2}e^{-2\psi}a_\pm^{2}f_{\mu\nu,y}
	+\mu\,a_\pm^{-1/p^2}\partial_\mu\phi_\pm\partial_\nu\phi_\pm
	\right]     \ .
\end{eqnarray}
Thus the trace part of extrinsic curvature on each brane is obtained as
\begin{eqnarray}
{\cal K}_\pm&=&g^{\mu\nu}_\pm{\cal K}^\pm_{\mu\nu}\nonumber\\
	&=&n_y\left[
	4\mu\,e^{-2\psi}a_\pm^{-1/p^2}+a_\pm^{-2}\square\phi_\pm
	+\mu\,a_\pm^{-2-1/p^2}(\partial\phi_\pm)^2
	+2a_\pm^{-2}(\partial_\alpha\phi)(\partial^\alpha\psi)
	+\frac{1}{6\mu}a_\pm^{-2+1/p^2}\left(R-3p^2(\partial\psi)^2\right)
	\right.\nonumber\\
&&	\left.\qquad
	+\frac{\kappa}{\sqrt{3}}e^{-2\psi}\left\{
	\frac{3(2p^2-1)\mu}{p^3}a_\pm^{-2+1/p^2}\beta
	+\frac{4\mu}{p}a_\pm^{-2/p^2}C_1
	+\frac{2(4p^2-1)\mu}{p^3}a_\pm^{-4}C_2
	\right\}\right] \ .
\end{eqnarray}


\end{document}